\documentclass[twocolumn]{openjournal}

\usepackage{xcolor}
\usepackage{textgreek}
\usepackage[utf8]{inputenc}
\usepackage[english]{babel}

\usepackage{hyperref}
\hypersetup{
    unicode, 
    colorlinks=true,
    linkcolor=linkcolor,
    citecolor=linkcolor,
    filecolor=linkcolor,
    urlcolor=linkcolor,
}
\usepackage{color,colortbl}
\definecolor{linkcolor}{rgb}{0.0,0.3,0.5}
\usepackage{tensind}
\tensordelimiter{?}
\DeclareGraphicsExtensions{.bmp,.png,.jpg,.pdf}
\usepackage{verbatim}
\usepackage[normalem]{ulem}
\usepackage{orcidlink}
\usepackage{soul}

\begin{document}

\title{Modeling Carbon-Based Planets with MAGRATHEA}

\author{Robert Royer III\orcidlink{0000-0002-0900-0192}}
\email{royer@unlv.nevada.edu}
\affiliation{University of Nevada, Las Vegas}
\affiliation{Nevada Center for Astrophysics}

\author{David R. Rice\orcidlink{0000-0001-6009-8685}}
\affiliation{Department of Astronomy, University of Wisconsin–Madison}
\email{drice6@wisc.edu}

\author{Jason H. Steffen\orcidlink{0000-0003-2202-3847}}
\affiliation{University of Nevada, Las Vegas}
\affiliation{Nevada Center for Astrophysics}

\begin{abstract}

A planet's interior structure not only influences its habitability, but it also provides information about the planet's formation and history.  In the quest to characterize the variety of planetary discoveries, the most common practice is to rely on our understanding of the Earth's differentiated interior structure, with an iron core and silicate mantle as a starting point.  However, recent work by \citet{cody25} shows that these assumptions may not be true for all planetary systems as the composition of the condensed material can change drastically towards a carbon mantle instead of a silicate one, once the C/O ratio of the system approaches and exceeds $\sim 0.9$.  In an effort to more fully understand the range of possible planetary bodies, we added the ability to model carbon-based mantles to our open-source planetary interior solver, MAGRATHEA.  The newly added carbon mantle includes three relevant allotropes of carbon: graphite, diamond, and BC8 (body centered cubic with eight atoms per cell).  
\end{abstract}

\section{Introduction}
Our solar system is one of the largest and most varied planetary systems known. It hosts several terrestrial planets as well as gas and ice giants.  This diversity does not merely extend to their masses and radii, but is also seen in the different interior structures.  For example, the Earth's core contains approximately 30\% of its mass, while Mercury's core contains nearly 70\% of the planet mass \citep{Hauck2013}.  This diversity shows that the inner solar system complicates the peas-in-a-pod trend seen in other, more compact, multiplanet systems (\citet{lammers23}, \citet{weiss22}).  These differences suggest that the planets of our inner solar system may be the result of different formation histories.

The core mass fraction (CMF) of rocky bodies in the inner solar system ranges from approximately 20\% to 70\% \citep{Szurgot15}.  The differences that we see in the interior structure of planets in the inner solar system have a large impact on habitability, affecting properties such as the temperature profile, volatile abundance, and the presence of a magnetic field (\citet{Lammer09}, \citet{Lichtenberg23}). One common feature shared across the inner solar system, though, is the presence of an iron core and a silicate mantle.  

Taking the Earth's differentiated structure as a blueprint, many studies developed mass-radius relationships to characterize the growing number of observed exoplanets (i.e. \citet{Zeng2016}, \citet{Dorn2017}).  A significant limiting factor in relying solely on mass-radius relationships is the degeneracies present in solutions for a given mass and radius \citep{Rogers2010}, most notably when we can infer a significant atmosphere or hydrosphere.

This has led to debate regarding the compositions of planets above two Earth masses and uncertainty whether these planets should be classified as super-Earths or sub-Neptunes.  One way this issue has been addressed is by studying multi-planet exoplanet systems, and invoking a uniform composition model \citep{Agol2020} across a host star's planetary system regardless of each planets bulk density.

Another proposed method to constrain exoplanet compositions is to look at the characteristics of the host stars.  Studies such as \citet{Rogers2010} and \citet{Dorn2017} suggest that the overall planetary composition may be coupled to stellar abundances.  This characteristic has been identified in several small exoplanets as well (i.e. \citet{DressingCharb15}) and studies such as \citet{Brugger2017} have proposed this as a constraint for interior modeling. 

As the chemical evolution of material within our galaxy continues, it will be shaped more heavily by the products of low mass stars, owing to their large population density and long stellar lifetimes.  One important chemical that is released with the death of low mass stars is carbon \citep{Johnson2019}.  Recent work by \citet{steffen25} shows that as the galactic age increases, we can expect noticeable differences in the bulk composition and relative size of newly formed planets.  We can extrapolate this to the carbon and silicon regimes of planetary mantles as well.

\citet{cody25} shows that there are three regimes of protoplanetary disks around stars with different carbon-to-oxygen (C/O) ratios: less than 0.6, between 0.6 and 0.9, and greater than 0.9.  Protoplanetary disks are silicon-dominated when the C/O ratio of the host star is less than 0.6.  In this regime, the solid surface density of carbon is exceedingly low and the majority of mass in the inner disk is composed of silicates.  For stars with a C/O ratio between 0.6 and 0.9, the disks have a lower amount of silicate material, and carbon starts to become significant.  Stars with a C/O ratio above 0.9 are in the carbon-dominant regime, with up to 80\% of the disk's condensed mass made of carbon \citep{cody25}.

The disk from which our solar system formed was in the lowest C/O regime with ratios below 0.6 (the solar C/O is 0.5) and had an inner disk dominated by silicate materials.  Disks forming in the highest C/O regime are likely to have disks dominated by carbon material, out to as far as 6 Au \citep{cody25}.  Taking these two observations together, one may infer that as stars form later in the chemical enrichment cycle, or having a high concentration of carbon due to the natural variations in the galactic material, their disks will begin to be dominated more heavily by carbon.  Planets that form in these carbon enriched disks will be more likely to have carbon dominated mantles.   

Even with an average disk C/O ratio lower than 0.9, regions of the protoplanetary disk can be locally enriched in carbon material as well.  One of these regions that has recently received renewed interest is the area between the carbon and water ice lines in the protoplanetary disk.  In this region, carbon-bearing molecules are able to condense into the solid phase, while the disk is still too hot to form water ices.  Planets that form in this region are thought to be highly enriched in carbon, and has lead to the recent interest in soot planets \citep{Li26}.

Another region of the protoplanetary disk that is likely to favor the formation of carbon dominated bodies is the circumplanetary disk that forms around giant planets in the outer regions of the protoplanetary disk.  Recent work by \citet{Boss26} shows that these circumplanetary disks can have supersolar C/O ratios, which would be critical for the formation of carbon-dominant moons.

Carbon-based planets have been the subject of studies by other authors.  \citet{Seager2007} modeled carbon-dominant planets in their study of planetary mass-radius relationships.  They found that the mass-radius relationship for planets with either a graphite or silicon carbide mantle intersect with those of silicate planets and water worlds.  Following the discovery of 55 Cancri e, \citet{Madhusudhan12} modeled both carbon and silicon carbide planets to match the observational constraints.  They found that a carbon-dominant mantle can plausibly match observations without the need for the significant hydrosphere needed to match 55 Cancri e's bulk density when modeling with a silicate mantle.  \citet{Wilson14} investigated the mass-radius relationships for silicon carbide planets, finding two new forms of high-pressure silicon carbide.  One particularly interesting approach of their work was to simulate planets with a diamond mantle and silicon carbide core.  They found that the density of planets with a silicon carbide composition to fall between the densities of pure carbon and pure silicate planets.  

Modeling the interior structure of these diverse planetary bodies is necessary for our understanding of the population of exoplanets in our galaxy, as is putting these tools into the hands of the research community.  Understanding the structure of these planets is important for observational campaigns as these models will allow for more meaningful mass-radius relationships to be identified, and will allow us to more meaningfully gauge the prevalence of carbon-dominant exoplanets. To achieve these aims, we have implemented equations of state for three astrophysically relevant carbon allotropes into our publicly available one-dimensional interior structure solver, MAGRATHEA, allowing for more meaningful simulations of carbon-rich, rocky planets.  As we continue to improve upon MAGRATHEA, we plan to incorporate these new carbon equations of state into a mixing algorithm.  Incorporating mixing into future MAGRATHEA builds will allow users to simulate carbon and silicon carbide planets by defining relative amounts of carbon and silicon in the mantle.

In section \ref{sec:software} we discuss the MAGRATHEA software.  In section \ref{sec:EoS} we list the equations of state for each carbon phase implemented.  In section \ref{sec:conclusion} we discuss the results of our work, and possible next steps in refining the capabilities of MAGRATHEA.

\section{Software} \label{sec:software}

MAGRATHEA is an open source, spherically symmetrical interior structure solver for planetary interiors developed by \citet{Huang22}, written in C++.  MAGRATHEA supports interior structure simulations for planets of up to four layers: core, mantle, hydrosphere, and atmosphere.  The code takes user defined masses for each differentiated layer and calculates the radius, density, pressure, and temperature using equations of hydrostatic equilibrium \citep{Huang22}.  It uses the Runge-Kutta-Fehlberg and outside-in, shoot-to-fit methods to solve the ordinary differential equations for mass continuity, hydrostatic equilibrium, and temperature gradient \citep{Huang22}.  

\begin{figure}
        \centering
        \includegraphics[width=\columnwidth]{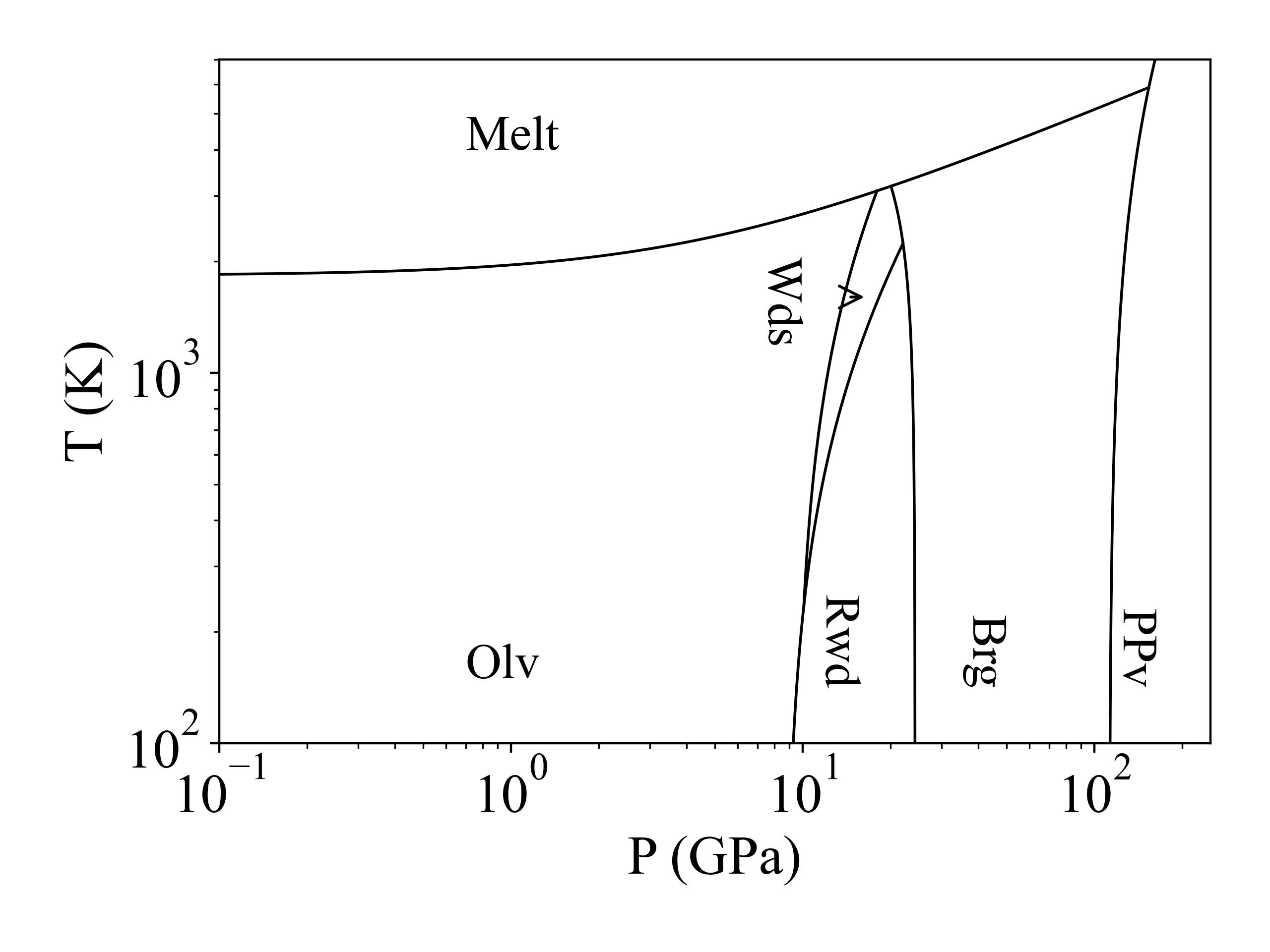}
        \label{fig:si_phase}
    \hfill
        \centering
        \includegraphics[width=\columnwidth]{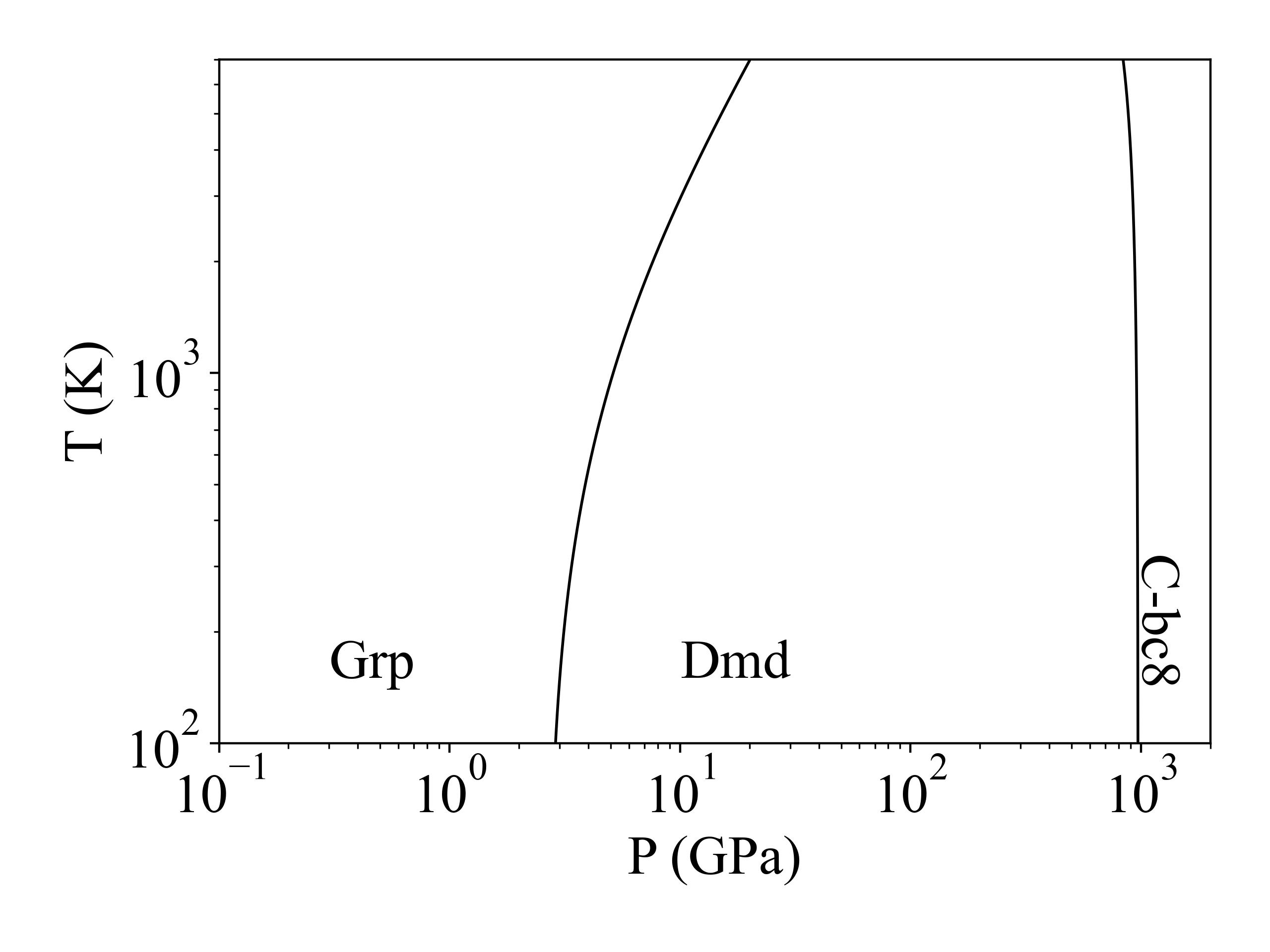}
        \label{fig:c_phase}
    \hfill
    \caption{Phase diagrams for the default silicon mantle (top) and newly implemented carbon (bottom).  Pressure is in the units GPa, and temperature in K}
    \label{fig:phases}
\end{figure}

Each differentiated layer simulated in MAGRATHEA has its own accompanying phase diagram (see \ref{fig:phases}).  Using the calculated pressure and temperature at each point, the phase diagram is used to determine the appropriate allotrope and density of each material.  The default phase diagrams for MAGRATHEA assume an iron core, a silicon-dominated mantle, a pure water/ice hydrosphere, and ideal gas atmosphere.  Users can simulate a continuous thermal profile between each layer, or invoke temperature discontinuities between any or all layers.  

To run MAGRATHEA, the user first enters the desired parameters into the appropriate configuration file (e.g., the mass of the planet layers, the composition of the layers, the temperature at the surface, temperature jumps between the layers, etc.).  
The software we developed here gives Carbon as an option for the mantle composition.  To use the new carbon phase diagram, the "mantle\_phasedgm" variable in the input configuration file must be changed from "Si\_default" to "C\_simple" in the configuration file of the desired input mode (i.e. mode0.cfg).

\section{Equations of State} \label{sec:EoS}

Each phase on the phase diagram for a given planetary layer has an associated equation of state that determines the density of the material as a function of the thermodynamic conditions such as temperature and pressure.  For the carbon layer, we include four new equations of state (EoS) for three astrophysically relevant phases: two equations for graphite, diamond, and the theoretical BC8 (body centered cubic structure with eight atoms per unit cell).  One of the graphite EoSs is isothermal, while the second allows for the modeling of thermal pressure (which is the default in our new carbon phase diagram).

\begin{figure}
        \centering
        \includegraphics[width=\columnwidth]{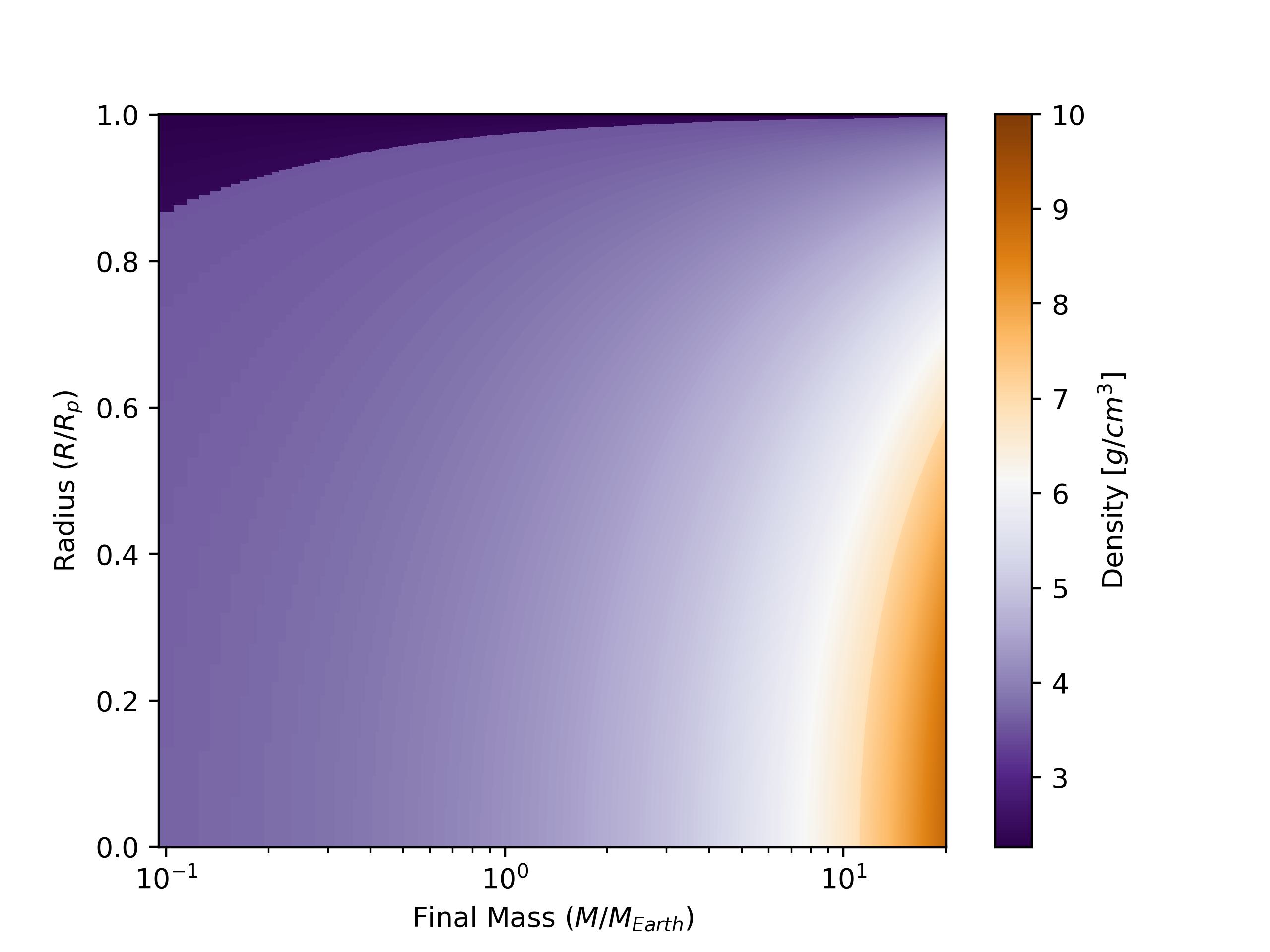}
    \caption{Figure showing the structure of pure carbon planets ranging from 0.1 to 20 Earth masses with a surface temperature of 300K.  Color bar indicates density in g/cm$^{3}$.}
    \label{fig:pure_C}
\end{figure}

\begin{figure}
    \centering
    \includegraphics[width=\columnwidth]{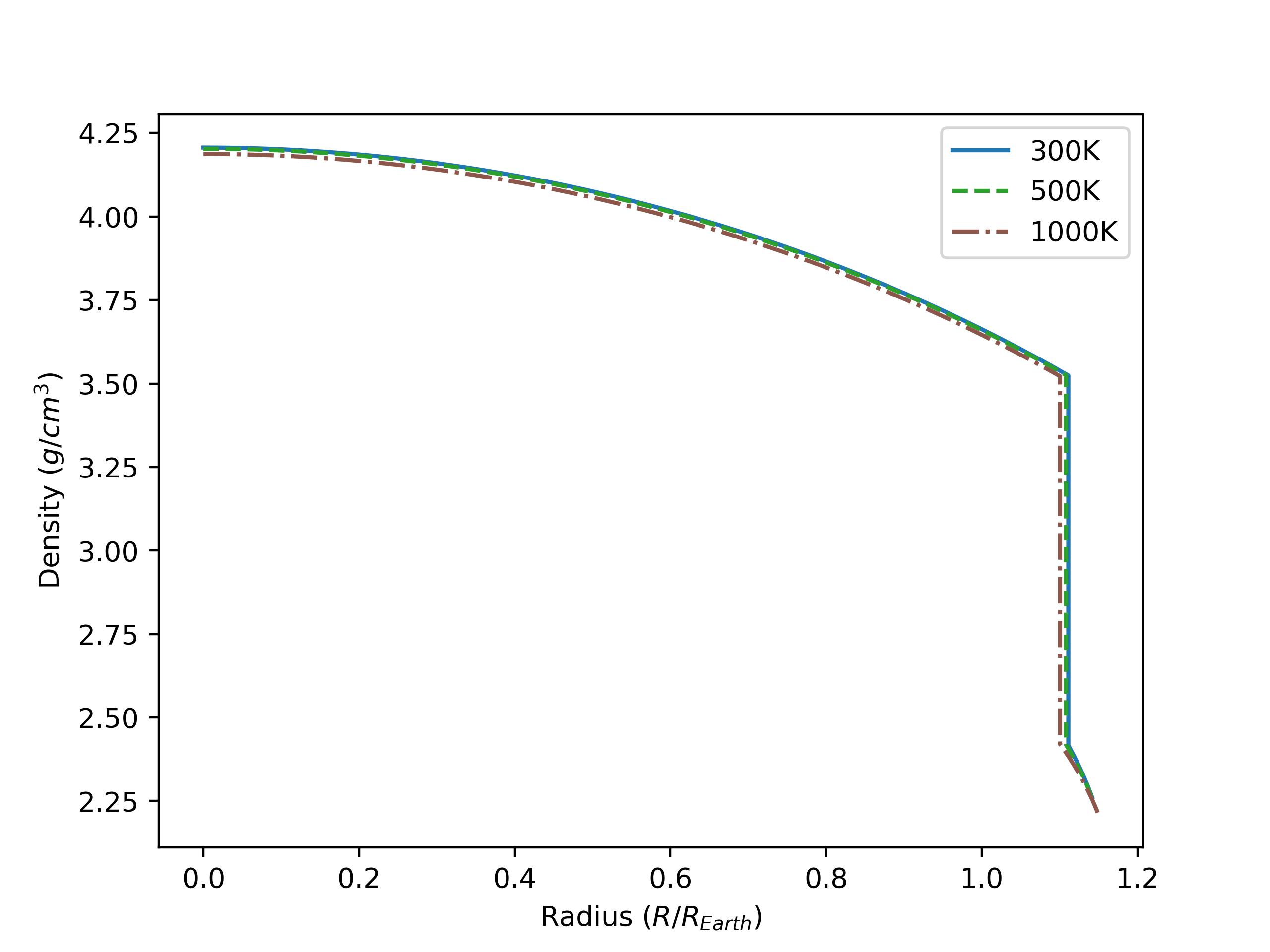}
    \label{fig:pure_C_density}
    \caption{Density vs radius plot for pure carbon planets with different surface temperatures.  Planets contain one Earth mass with no iron core.  The solid blue line corresponds to a surface temperature of 300K.  The dashed green line represents a surface temperature of 500K, and the dash-dotted brown line represents a surface temperature of 1000K.}
\end{figure}
    
\subsection{Graphite}

Graphite represents the topmost layer in our interior model, and as such has the lowest density of the carbon phases.  We include two EoSs to model the graphite layers in MAGRATHEA.  The first implemented graphite EoS is taken from \citet{Seager2007} and is an isothermal third-order, Birch-Murnagham EoS of the form:  
\begin{equation}
    P = \frac{3}{2}K_{0}(\eta^{7/3}-\eta^{5/3})[1+\frac{3}{4}(K_{0}'-4)(\eta^{2/3}-1]
    \label{eq:BME3}
\end{equation}

where $\eta = V_{0}/V$, $K$ is the isothermal bulk modulus, and $K'$ is the first derivative of the bulk modulus with respect to pressure.

We have also implemented a high-temperature Vinet EoS with Debye model thermal parameters taken from \citet{Lowitzer} of the form: 
\begin{equation}
    P = 3K_{0}\eta^{2/3}(1-\eta^{-1/3})exp[\frac{3}{2}(K_{0}'-1)(1-\eta^{-1/3})].
    \label{eq:vinet}
\end{equation}
\citet{Lowitzer} provides two Vinet EoSs for the graphite phase with the only difference being the pressure derivative of the bulk modulus.  We have chosen to implement the EoS with a lower pressure derivative ($K' = 9$) as it matches more closely to the empirical data \citep{Lowitzer}.  They also use the Mie-Gruneisen-Debye framework to model the thermal effect on the reference volume, using the modification:
\begin{equation}
    V_{0T} = V_{0}exp[\int_{T_{0}}^{T}(\alpha_{0}+\alpha_{1}T)dT]
\end{equation}

We take our graphite/diamond phase transition from \citet{Kennedy} and is represented by equation \ref{eq:1}.  

\begin{equation}
    P = 1.949 + \frac{T(K)}{400} [GPa]
    \label{eq:1}
\end{equation}

\begin{table}
\begin{tabular}{|ll|l|l|}
\hline
\multicolumn{2}{|l|}{} & \textbf{Seager} & \textbf{Lowitzer} \\ \hline
\multicolumn{1}{|l|}{$\mathrm{V}_0$} & $\mathrm{cm}^3/\mathrm{mol}$ & 5.33822 & 5.31153 \\ \hline
\multicolumn{1}{|l|}{$\mathrm{K}_0$} & GPa & 33.8 & 38 \\ \hline
\multicolumn{1}{|l|}{$\mathrm{K}'_0$} &  & 8.9 & 9 \\ \hline
\multicolumn{1}{|l|}{$\mathrm{M}_{\mathrm{mol}}$} & g/mol & 12.011 & 12.011 \\ \hline
\multicolumn{1}{|l|}{n} &  &  & 1 \\ \hline
\multicolumn{1}{|l|}{Z} &  &  & 6 \\ \hline
\multicolumn{1}{|l|}{$\mathrm{T}_0$} &  &  & 300 \\ \hline
\multicolumn{1}{|l|}{$\alpha_0$} & $10^{-6}/\textrm{K}$ &  & 32 \\ \hline
\multicolumn{1}{|l|}{$\mathrm{\alpha_1}$} & $10^{-6}/\textrm{K}^{2}$ &  & 0 \\ \hline
\multicolumn{1}{|l|}{$\mathrm{C}_\mathrm{p0}$} & $10^{7} \textrm{erg}/\textrm{g K}$ &  & 0.714345 \\ \hline
\end{tabular}
\caption{Parameters implemented for both implemented forms of graphite.}
\end{table}

\subsection{Diamond}

Our diamond equation of state is taken from \citet{benedict14} and is a Vinet EoS with a modified Debye thermal profile to describe its complicated thermal behavior.  \citep{correa08} describe in detail the complicated nature of diamond and BC8 due to their double peaked phonon density of states.  This arises from the transverse acoustic waves being ``separated in energy from the rest of the phonon modes at high pressures'' \citep{correa08}.  To address this, \citet{benedict14} implements a double-Debye model to determine the Debye temperature for this phase.  While this process shows good agreement with their data, its form does not agree with the form expected by MAGRATHEA.  Three systems ('a', 'b', and '1') are included in their double-Debye model accounting for both the hot and cold portions of the diamond thermal profile, as well as an intermediary term corresponding to the transition between the hot and cold regimes.

We made significant effort to reduce the double-Debye model to the single Debye model, with unsatisfactory results.  Noting that there is less than a 10\% difference in the densities between the two thermal extremes, and that the intermediary term is necessarily bound between these two extremes, we chose to implement only the intermediary, '1', values into our phase diagram as a single-Debye model of the form:
\begin{equation}
    P_{TH} = \frac{\gamma}{V}E_{TH}+\frac{3nR}{2V}e_{0}x^ggT^2
\end{equation}
where
\begin{equation}
    x=\frac{V}{V_0}, \gamma = \gamma_{\infty}+(\gamma_0-\gamma_\infty)x^\beta,
\end{equation}
\begin{equation}
     \Theta = \Theta_0x^(-\gamma_\infty)exp[\frac{\gamma_0-\gamma_\infty}{\beta}(1-x^\beta],
\end{equation}
\begin{equation}
    z = \frac{\Theta}{T},E_{TH}=3nRTD_3(z).
\end{equation}
$D_3$ is the third-order Debye function. $\Theta_0,\gamma_\infty, \gamma_0, \beta, e_0, and g$ are fitting parameters for the Debye model \citep{Huang22}.  

\begin{figure}
        \centering
        \includegraphics[width=\columnwidth]{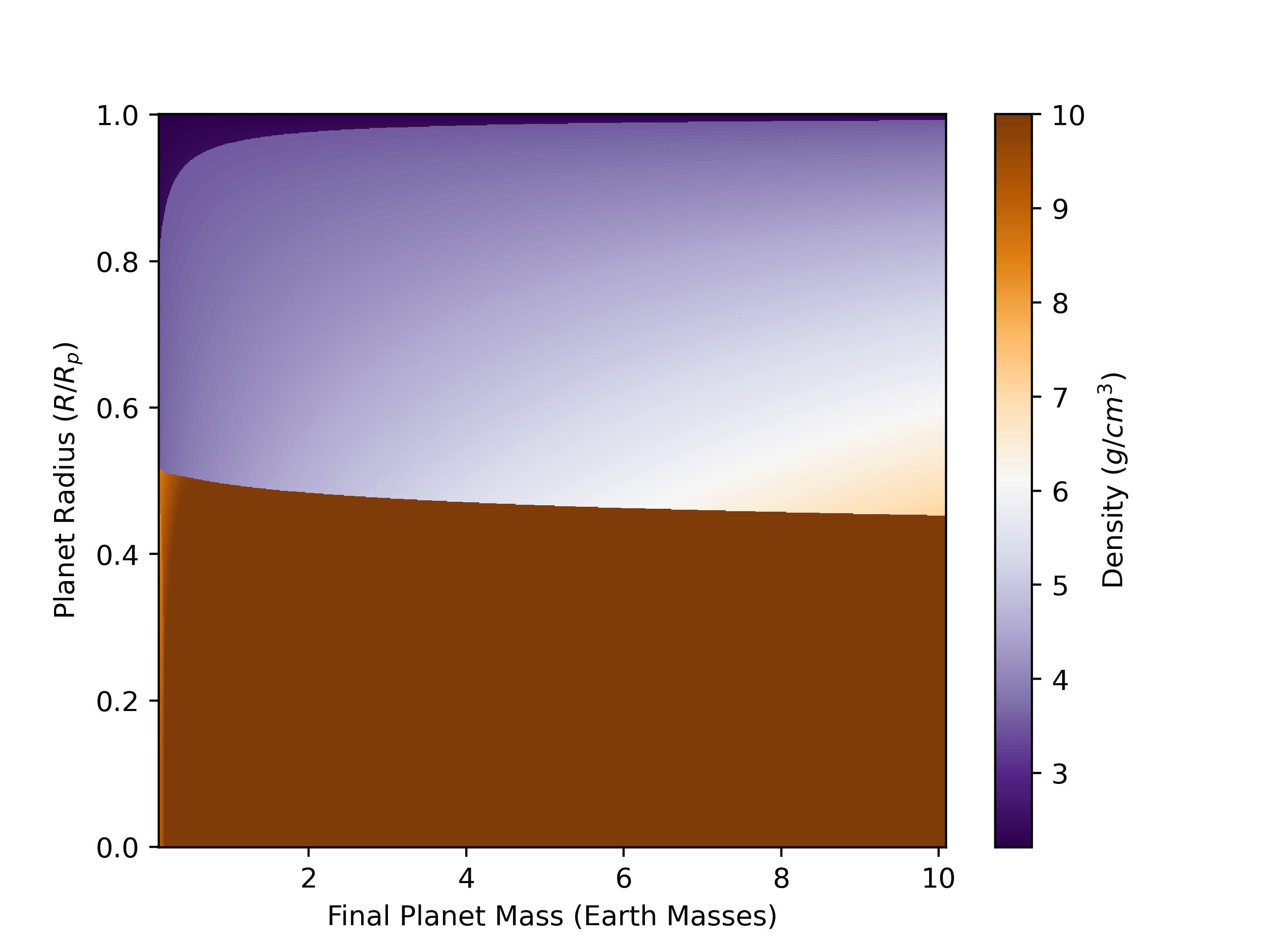}
        \label{fig:a}
    \hfill
        \centering
        \includegraphics[width=\columnwidth]{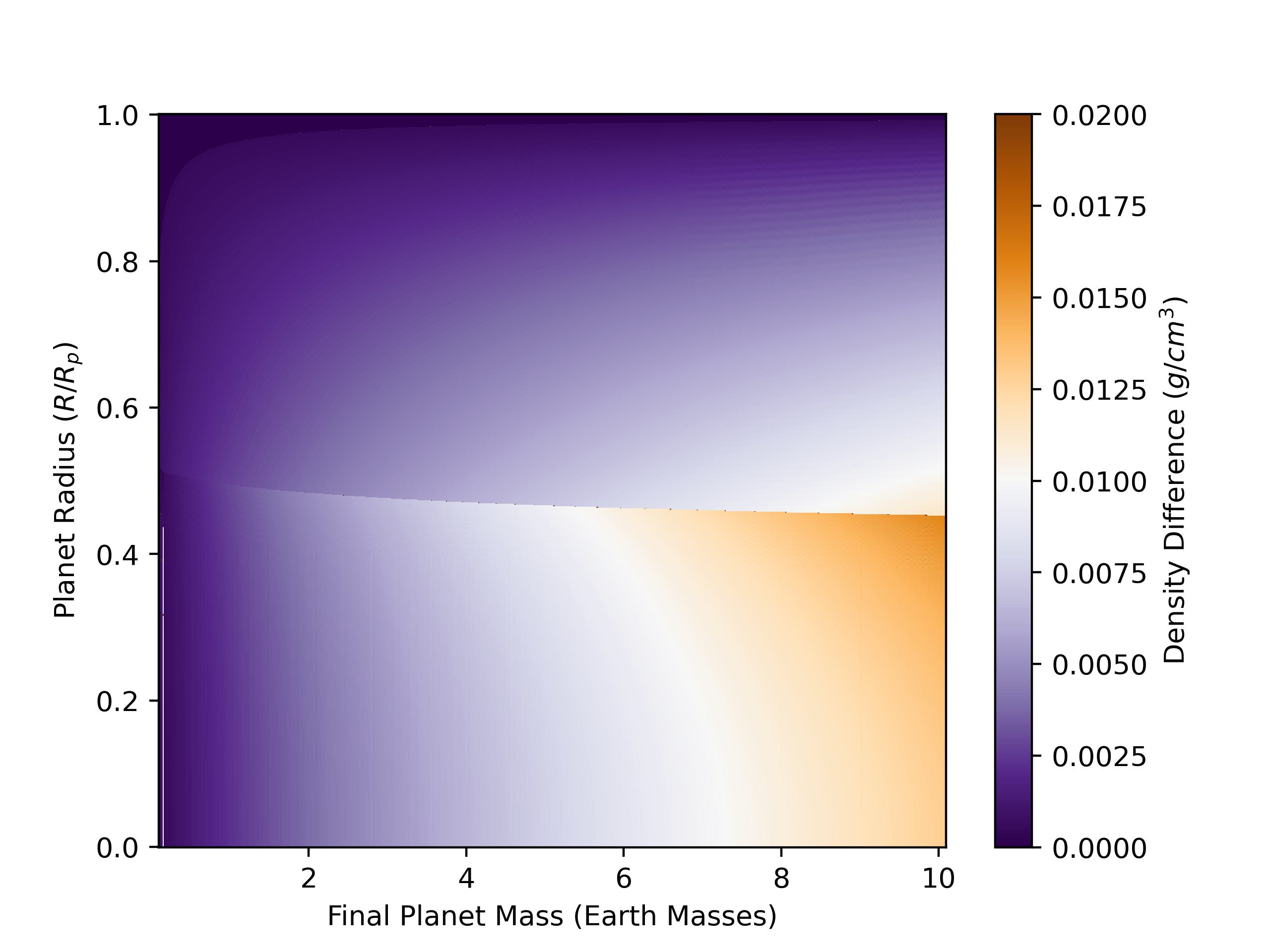}
        \label{fig:b}
    \hfill
        \centering
        \includegraphics[width=\columnwidth]{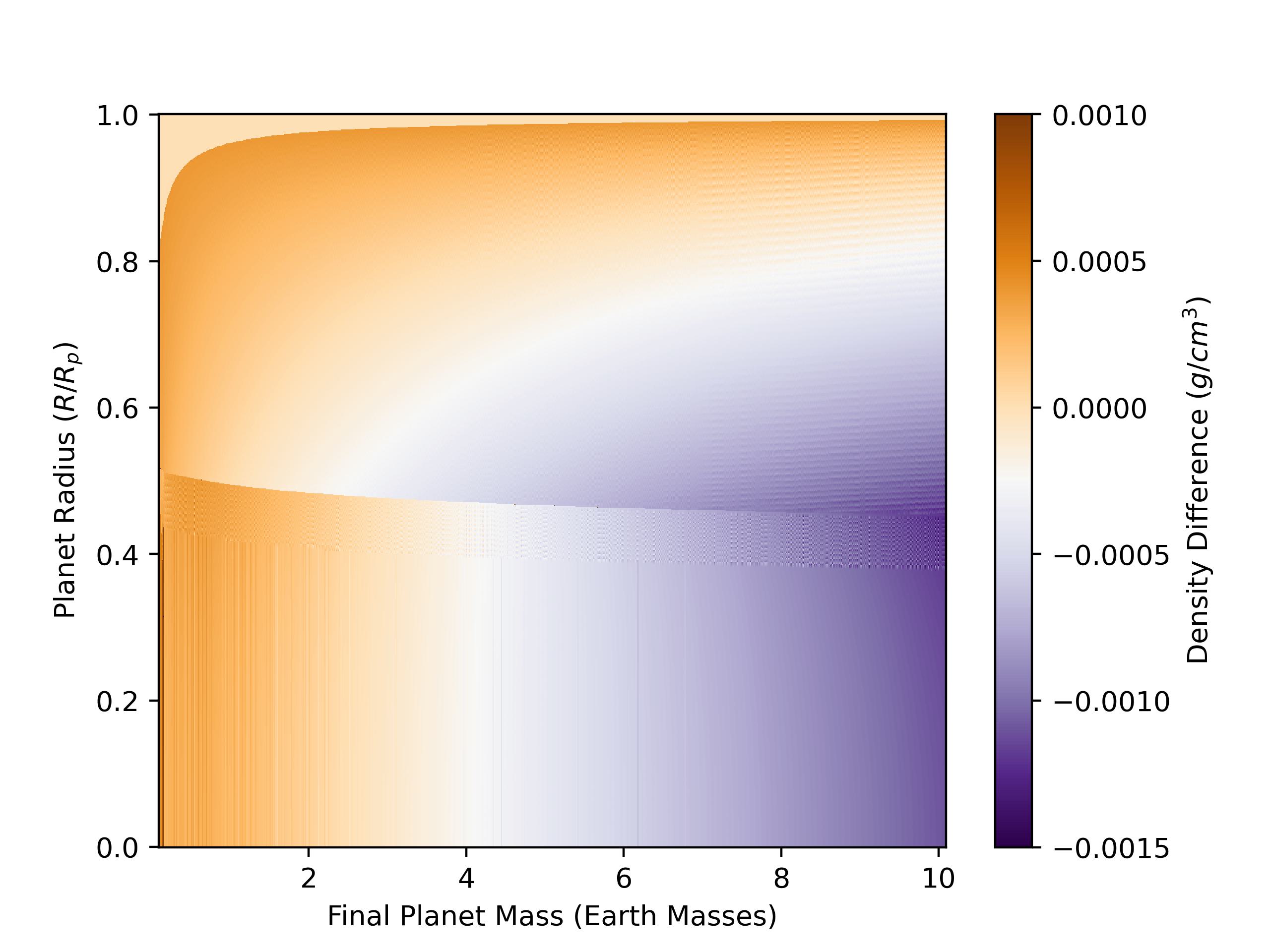}
        \label{fig:1}

    \caption{Internal structure for a planet with a graphite/diamond mantle.  Top panel uses the '1' terms from \citet{benedict14}.  Middle panel shows the difference in density between '1' terms and 'a' terms.  Lower panel shows the same for '1' terms and 'b' terms. '1' terms are implemented into MAGRATHEA, as they are the most intermediary of the three terms.}
    \label{fig:diamond}
\end{figure}

We tested each of the three terms independently as shown in \ref{fig:diamond}, and compared against each other to show that the simplifications to the diamond thermal profile are reasonable.  Each of the three terms are listed in \ref{tab:comp}.  We also considered several other works (i.e. \citet{correa08}, \citet{swift22}) for implementation into MAGRATHEA but discounted those due to implementation challenges arising from the necessary modifications to the EoS to fit the complicated thermal profile, as mentioned above.  

\begin{table}
\begin{tabular}{|ll|l|l|}
\hline
\multicolumn{2}{|l|}{ \textbf{Parameter/Units} } & \textbf{Diamond} & \textbf{BC8} \\ \hline
\multicolumn{1}{|l|}{$\mathrm{V}_{0}$} & $\mathrm{cm^3/mol}$ & 3.43467 & 3.75902 \\ \hline
\multicolumn{1}{|l|}{$\mathrm{K}_{0}$} & GPa & 432.4 & 221.2 \\ \hline
\multicolumn{1}{|l|}{$\mathrm{K}'_{0}$} &  & 3.793 & 4.697 \\ \hline
\multicolumn{1}{|l|}{$\mathrm{M}_{mol}$} & g/mol & 12.011 & 12.011 \\ \hline
\multicolumn{1}{|l|}{$\mathrm{\theta_0}$} & K & 1887.8 & 2800.6 \\ \hline
\multicolumn{1}{|l|}{$\mathrm{\gamma_0}$} &  & 0.5836 & 0.561 \\ \hline
\multicolumn{1}{|l|}{$\mathrm{\beta}$} &  & 1 & 1 \\ \hline
\multicolumn{1}{|l|}{$\mathrm{\gamma_\infty}$} &  & 0.499 & 0.449 \\ \hline
\multicolumn{1}{|l|}{n} &  & 1 & 1 \\ \hline
\multicolumn{1}{|l|}{Z} &  & 6 & 6 \\ \hline
\end{tabular}
\label{tab:diam}
\caption{Implemented parameters for diamond and BC8 phases of carbon, both containing the '1' terms taken from the double-Debye model in \citet{benedict14}.  See \ref{tab:comp} for a comparison of terms for each state.}
\end{table}

\begin{table}
\begin{tabular}{|l|l|l|l|}
\hline
\textbf{Diamond} & \textbf{a} & \textbf{b} & \textbf{1} \\ \hline
\textbf{$\mathrm{\theta_0}$} & 1887.8 & 1887.8 & 1887.8 \\ \hline
\textbf{$\mathrm{\gamma_0}$} & 0.597 & 0.597 & 0.5836 \\ \hline
\textbf{$\mathrm{\gamma_\infty}$} & 0.913 & 0.429 & 0.499 \\ \hline
\textbf{BC8} & \textbf{a} & \textbf{b} & \textbf{1} \\ \hline
\textbf{$\mathrm{\theta_0}$} & 1961.9 & 3176.3 & 2800.6 \\ \hline
\textbf{$\mathrm{\gamma_0}$} & 0.0 & 0.688 & 0.561 \\ \hline
\textbf{$\mathrm{\gamma_\infty}$} & 0.0 & 0.532 & 0.449 \\ \hline
\end{tabular}
\label{tab:comp}
\caption{Comparison of 'a', 'b', and '1' terms for diamond and BC8 found in \citet{benedict14}.  '1' terms are implemented into MAGRATHEA, as they represent and intermediary value between 'a' and 'b' terms.}
\end{table}

\subsection{BC8}

Our BC8 EoS is also taken from \citet{benedict14}.  Similar to their treatment of the diamond thermal profile, they again use a double-Debye method to characterize the thermal properties of BC8.  For the same reasons listed in the above section we have chosen to implement the '1' terms for the BC8 phase.  In \ref{fig:pure_C} we see that the BC8 phase of carbon does not appear until the planet mass is above 10 Earth masses, implying that such a phase would not be relevant in the mantles of most rocky exoplanets, though, this phase may be relevant for larger carbon-based exoplanets, such as sub-Neptune or larger planets (i.e. soot planets).

\section{Carbon Earth}

To test the implementation of our phase diagram, we produced a series of simulated planets having the same core mass fraction (CMF) and mantle mass fraction (MMF) as Earth.  To model the core, we use the default core phase diagram for MAGRATHEA, consisting of the hexagonal close-packed (hcp) and liquid phases of iron.  Our mantle phase diagram uses the carbon phase diagram presented here (Figure \ref{fig:c_phase}).  

We ran simulations of planets ranging from 0.1 to 20.1 Earth masses with different surface and core-mantle boundary temperatures.  We used surface temperatures of 300K and 1,200K, corresponding to the ambient surface temperature of Earth and a temperature near Earth's mantle surface temperature, respectively.  We also simulated planets with two differing values for the temperature discontinuity at the core-mantle boundary (CMB), a continuous temperature profile (0K discontinuity) and a temperature discontinuity of 1,500K.  Taken together, this provides four different thermal profiles for testing.

When looking at simulated planets with one Earth-mass, planets simulated with a surface temperature of 1,200K and CMB temperature discontinuity of 1,500K have the largest radius, though all four Earth-mass planets range between 1.047 -- 1.058 Earth radii.  Planets with a surface temperature of 1,200K have the thickest graphite layer, at 0.046 Earth radii.  The magnitude of the CMB temperature discontinuity does not affect the thickness of the graphite layer, and has only minimal effects on the diamond layer.

The largest diamond layer is present in the 300K surface temperature planets with no CMB temperature discontinuity, with a thickness of 0.504 Earth radii.  In the case of a 1,200K surface temperature with 1,500K CMB temperature discontinuity, the thickness of this layer is reduced slightly to 0.503 Earth radii.  For a surface temperature of 1,200K the diamond layer has thicknesses of 0.491 and 0.489 for planets with no CMB temperature discontinuity and 1,500K, respectively.  The small differences we see in the thickness of the diamond layer show that the magnitude of the mantle temperature doesn't have a significant effect on the mantle's diamond layer.  At one Earth mass, none of the four thermal profiles had temperature and pressure conditions are sufficient to form BC8 layers of the carbon mantle.

\begin{figure}
        \centering
        \includegraphics[width=\columnwidth]{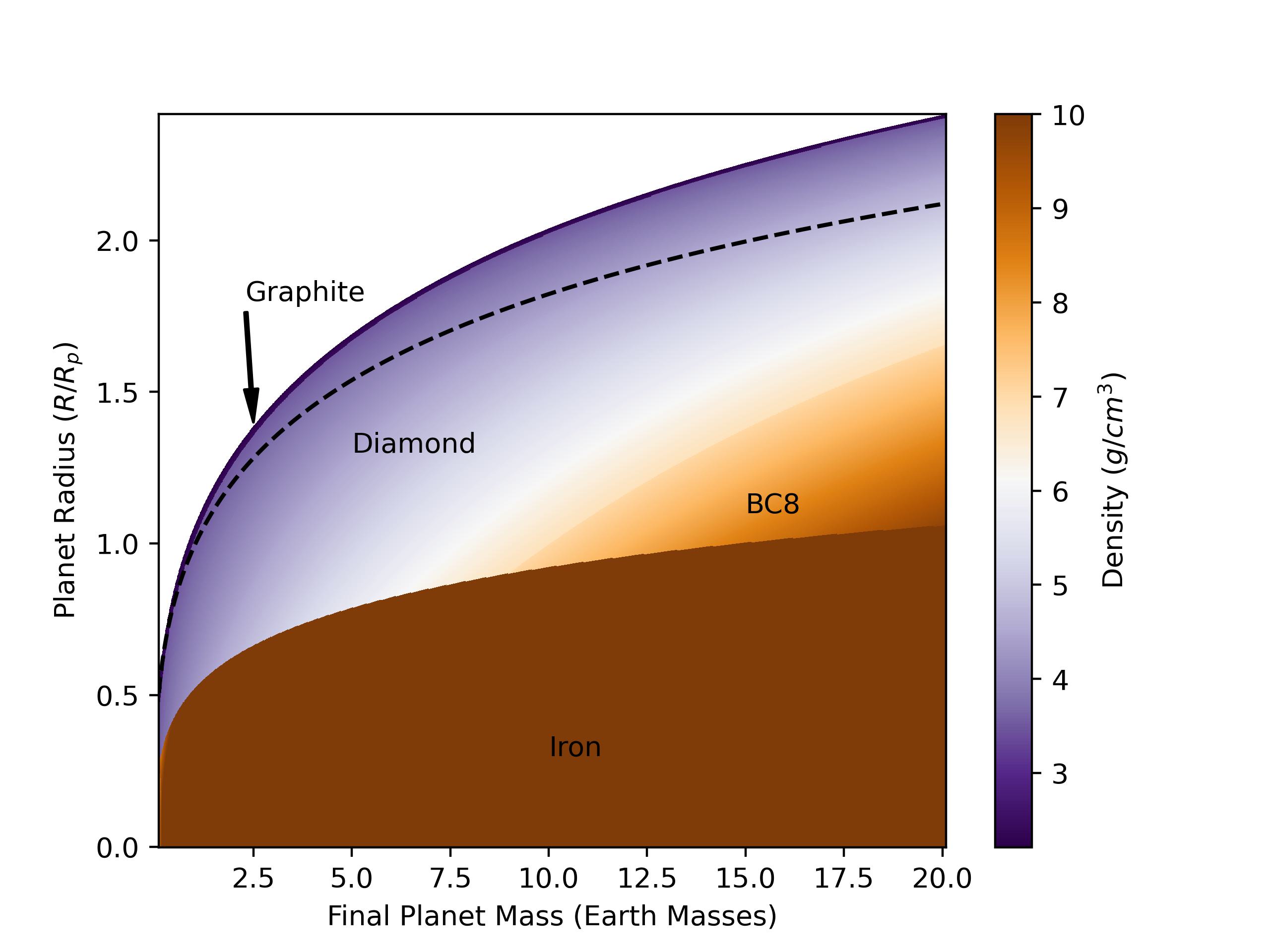}
        \label{fig:carbon unscaled}
    \caption{Carbon planets with an Earth-like CMF ranging from 0.1 to 20.1 Earth masses.  Dark purple, light purple, light orange, and dark orange correspond to graphite, diamond, BC8, and iron, respectively.  The dotted black line is the radius of silicon planets of the same mass.}
\label{fig:carbon comp}
\end{figure}

\begin{figure}
    \centering
    \includegraphics[width=\columnwidth]{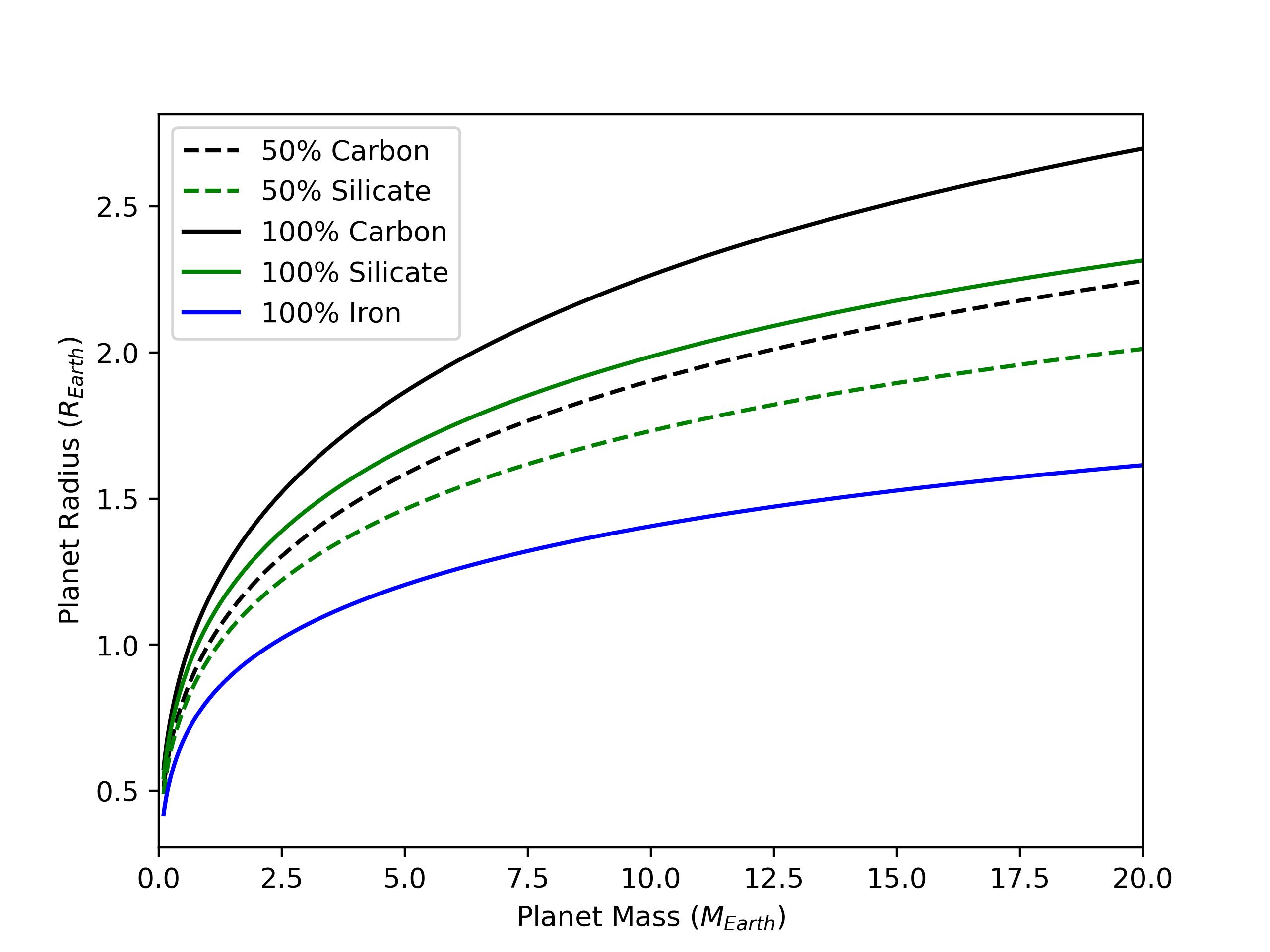}
    \label{fig:MR}
    \caption{Radius vs mass plots for planets ranging from 0.1 to 20 Earth masses.  Black lines correspond to carbon mantle planets.  Darker gray lines show silicate mantle planets.  Lighter gray lines correspond to pure iron planets.}
\end{figure}

\subsection{Carbon and Silicon Planet Comparison} \label{sec:comparison}

We compared the radius of planets with either a carbon or silicon mantle, using the default mantle phase diagram of MAGRATHEA (Figure \ref{fig:si_phase}).  The silicon phase diagram incorporates olivine, wadsleyite, and other phases of magnesium silicates (for additonal details see \citet{Huang22}).  We found that carbon planets are also larger than silicon-dominant planets of an equivalent mass.  Keeping an Earth-like CMF with a CMB temperature discontinuity of 1,500K and a surface temperature of 1,000K, we ran simulations ranging from 0.1 to 20.1 Earth masses with either a carbon or silicon based mantle. The radii of the carbon planets ranged from 0.535 Earth radii at 0.1 Earth masses to 2.41 Earth radii at 20.1 Earth masses, as seen in Figure \ref{fig:carbon comp}.  Silicon planets range from 0.511 to 2.12 Earth radii over this same mass range.  This equates to radius differences as large as 13.7\% for the most massive planets with the carbon planets being larger due to carbon being more stiff than silicates.  While carbon planets show a larger radius across all simulations, these radii are not sufficient to distinguish carbon planets from silicate planets in any but the most extreme cases. 

\begin{figure}
        \centering
        \includegraphics[width=\columnwidth]{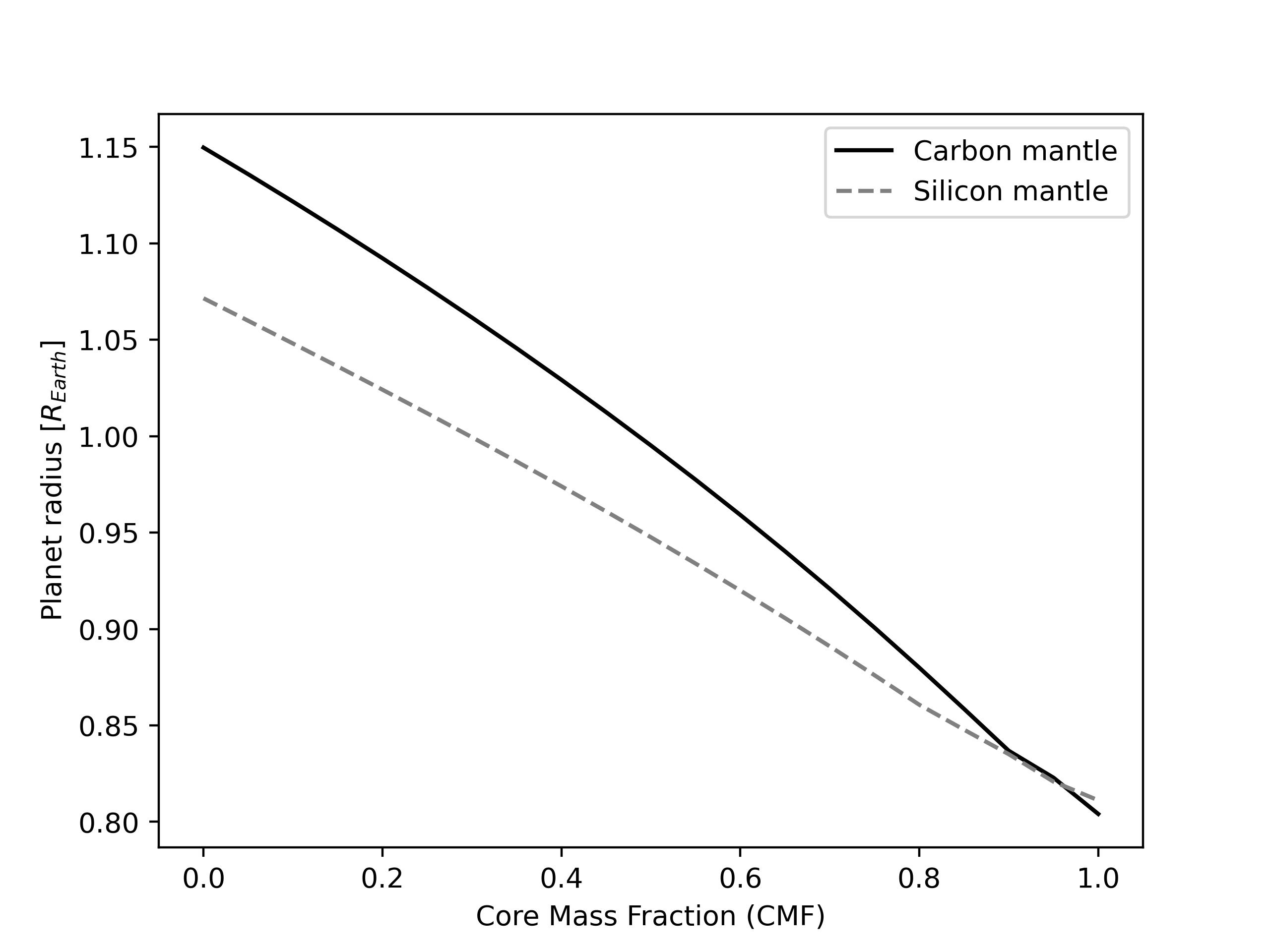}
\caption{Comparison of planet radius vs core mass fraction for planets with silicon and carbon mantles for Earth-mass planets.  The solid black and dashed gray lines correspond to planets with carbon and silicon mantles, respectively.}        
\label{fig:carbon CMF comp}
\end{figure}

We also see larger radii for carbon planets compared to silicate planets when we compare radius as a function of CMF (Figure \ref{fig:carbon CMF comp}).  Simulating both carbon and silicate Earth-mass planets ranging from 0 to 1 CMF, we see that carbon planets have a larger radius for every CMF less than 0.9.

\section{Comparisons with other works}

\begin{figure}
    \centering
    \includegraphics[width=\columnwidth]{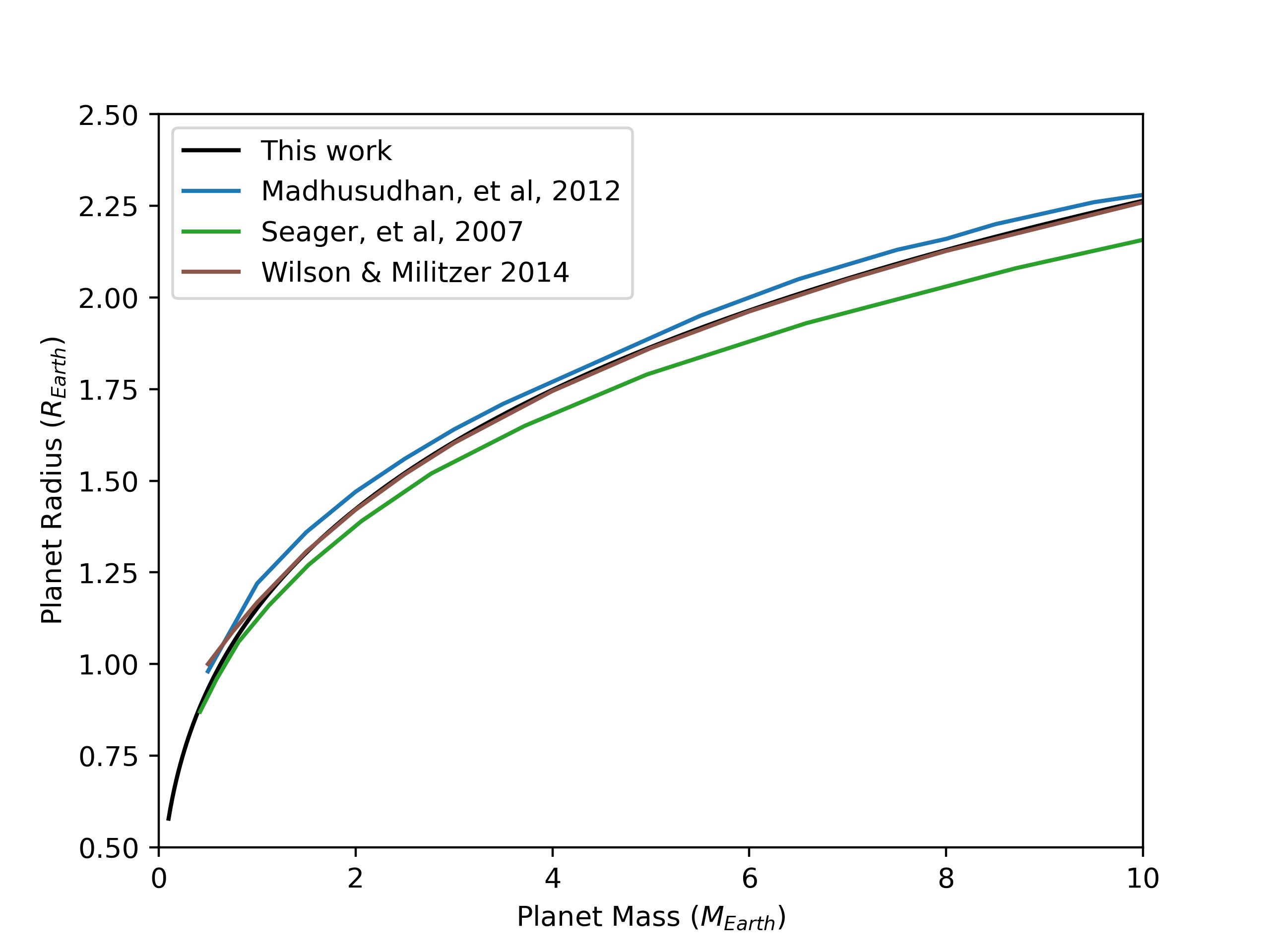}
    \label{fig:otherworks}
    \caption{Comparison of carbon planets generated with MAGRATHEA (black line) to selected previous works.  Blue line is taken from Figure 1 of \citet{Madhusudhan12}. Green line is extracted from Figure 9 of \citet{Seager2007}.  Brown line uses data from Figure 8 of \citet{Wilson14}.}
\end{figure}

Several previous works have investigated the properties of carbon planets.  We present our comparisons to three of those works with mass-radius curves generated with MAGRATHEA (Figure \ref{fig:otherworks}).  

\citet{Seager2007} looked at mass-radius relationships for several different planet compositions, including carbon and silicon carbide mantle planets.  The found that carbon planets with an iron core can have a similar density to that of planets with an iron core, silicate mantle, and significant water layer (water worlds).  The mass-radius relationship found in Figure 9 of their work shows a higher density than pure carbon planets generated by MAGRATHEA, but there are some differences between them.  \citet{Seager2007} planets have a CMF of 30\%, whereas the planets used in our mass-radius curve are composed entirely of carbon.  The higher CMF in \citet{Seager2007} is balanced by the fact that their mantle includes only graphite.  In our simulations the mantle includes the diamond and BC8 phases of carbon, which reduce the difference in overall planet radii.

\citet{Madhusudhan12} investigated carbon planets as a possible method to describe the observational measurements of 55 Cancri e without relying on a significant atmosphere.  They model planets with several different compositions including silicates, silicon carbide, carbon, and iron.  They found that carbon and silicon carbide planets were able to match the observational data for 55 Cancri e.  Similar to our work, they include graphite and diamond in their mantle layer, although they do not include the BC8 phase of carbon.  Comparing the results presented in Figure 1 of their work to our own, we see that their mass-radius relationships for a pure carbon planet are less dense than our own for planets between 1 and 10 Earth masses.

\citet{Wilson14} also investigated carbon and silicon carbide planets to describe observations of planets orbiting high C/O ratio stars, such as 55 Cancri e.  Although most of their work focuses on silicon carbide planets, they also simulate planets with a diamond/BC8 mantle, ignoring the low-pressure graphite phase.  They also disregard thermal effects as \citet{Seager2007} has shown that thermal effects have little effect on the planet composition (we also see this in our own simulations and can be seen in Figure \ref{fig:pure_C_density}).  The mass-radius curve for pure carbon planets presented in Figure 8 of their work is nearly identical to the mass-radius relationship presented here.

\section{Conclusion} \label{sec:conclusion}

We have added a new phase diagram to MAGRATHEA allowing users to simulate planets with a carbon mantle.  Our phase diagram for carbon planets contain three astrophysically relevant phases of carbon: graphite, diamond, and BC8.  We have implemented two phases of graphite from \citet{Seager2007} and \citet{Lowitzer}, allowing for user flexibility in the topmost layers of the carbon mantle.  Our diamond and BC8 phases are taken from \citet{benedict14}, and in both cases we use the '1' terms reflecting an intermediary between the high and low temperature extremes provided by the double-Debye model.

Simulating planets with Earth-like CMF/MMF ratios and masses, we find that carbon-based mantles result in planets approximately 5\% larger than the Earth for our four temperature profiles.  While it doesn't affect the thickness of the graphite surface layers, the temperature discontinuity at the core-mantle boundary slightly affects the diamond layer and overall planet radius, with slightly smaller diamond layers and larger planet radii for planets simulated with a 1,500K temperature discontinuity when compared to planets with no temperature discontinuity.  

Comparing carbon and silicon planets with equivalent masses, we see that carbon planets have a larger overall radius, up to 13.7\% larger for the most massive planets simulated.  

Our carbon phase diagram matches those proposed by previous works well.  Our mass-radius relationship aligns nearly identically to that put forth by \citet{Wilson14}, and show good alignment with those of \citet{Madhusudhan12} and \citet{Seager2007}.  

\begin{acknowledgments}
We would like to thank Chenliang Huang for developing the original MAGRATHEA software, and continuing to make updates and improvements to the code.
\end{acknowledgments}

\bibliographystyle{aasjournal}
\bibliography{references}{}

\end{document}